\documentclass[a4paper,10pt]{revtex4-2}
\usepackage{mathrsfs}
\usepackage{graphicx}
\usepackage{latexsym}
\usepackage{amsmath}
\usepackage{amssymb}
\usepackage{textcomp}
\usepackage{amsbsy}
\usepackage{graphics}
\usepackage{epstopdf}
\usepackage{color}

\allowdisplaybreaks[4]

\begin{document}

\tolerance=5000

\title{Holographic representation of the unified early and late universe via a viscous dark fluid}

\author{I.~Brevik,$^{1}$\,\thanks{iver.h.brevik@ntnu.no}
A.~V.~Timoshkin,$^{2,3}$\,\thanks{alex.timosh@rambler.ru}
}
 \affiliation{ $^{1)}$ Department of Energy and Process Engineering,
Norwegian University of Science and Technology, N-7491 Trondheim, Norway\\
$^{2)}$Institute of Scientific Research and Development, Tomsk State Pedagogical University (TSPU),  634061 Tomsk, Russia \\
$^{3)}$ Lab. for Theor. Cosmology, International Centre  of Gravity and Cosmos,  Tomsk State University of Control Systems and Radio Electronics
(TUSUR),   634050 Tomsk, Russia\\
}

\tolerance=5000

\begin{abstract}
In this article we apply the holographic principle for describing in
a unifying way the early and the late-time universe,when the general
equation of state contains a bulk viscosity. We use the idea of
a generalized cut-off holographic dark energy introduced by Nojiri
and Odintsov (2006,2017), and  study the evolution of the universe when
the equation of state has two power-law asymptotes. Analytical
expressions for  the infrared cut-offs in terms of the particle
horizon are obtained. The energy conservation laws as derived from
the holographic point of view, are given for various forms of the
thermodynamic parameters and for various forms of the bulk
viscosity. As a result, we obtain a unifying description of the
early and the late-time universe in the presence of a viscous
holographic fluid.

\end{abstract}


\maketitle

\section{Introduction}

One of the possible avenues for describing  the evolution of the
universe,  is the holographic approach, also called  the holographic
principle \cite{1,2,3}.  The holographic dark energy model can be
useful for the description of  quantum gravity. The holographic
principle has been  widely applied, both to the  early and to the
late-time universe \cite{4,5,6,7,8,9,10,11,12,13,14,15,16,17}.  The
generalized cut-off holographic dark energy (HDE) model was proposed
Nojiri and Odintsov \cite{1,2}, where infrared cut-off was
identified with a combination of various parameters: Hubble
constant, particle and future horizons, cosmological constant, and
universe life-time.  All known holographic dark energy models
represent just particular classes of the Nojiri-Odintsov HDE as  was
shown in [18-20]. Furthermore, the Nojiri-Odintsov HDE gives a
consistent basis for a novel generalized entropy while describing
the holographic universe, see [21]. The holographic theory of the
universe is well confirmed by  astronomical observations [22-27].
Different applications of the theory of dark energy were studied in
the reviews [28, 29].

One may ask:  is it possible to build a description both the
early-time and the late-time cosmic accelerating expansion of the
universe in a single cosmological model? The answer to it is
obtained in the article [30], where a unifying approach to
early-time and late-time universe based on phantom cosmology was
proposed. The cosmological model of a unified description of the
early and the late-time accelerated  universe, in terms of the van
der Waals equation of state for the cosmic viscous fluid,  was
investigated in Ref.~\cite{31}.

Considerations about   viscous fluid models started some time ago,
but their applications to the accelerating universe are rather
recent. Various aspects  of viscous cosmology models have been
discussed in
Refs.~\cite{32,33,34,35,36,37,38,39,40,41,42,43,44,45,46}. A unified
description of  dark energy and  dark matter in the standard
Friedmann-Robertson-Walker cosmology, in terms of  a single
dissipative unified dark fluid,  was proposed in Ref.~\cite{47},
where the dissipation was represented by a bulk viscosity with a
constant coefficient.

The structure of the article is as follows: In Section II we
introduce the main aspects of the holographic principle, following
the terminology from Ref.~\cite{1}.  In Section III we explain the
unified dark fluid cosmological model and discuss the structure of
the proposed equation of state. We apply various forms of the
thermodynamic parameter and bulk viscosity for the unified viscous
dark fluid model introduced in the work [48]. In Section IV we give
the conclusion.

\section{Generalized holographic description of the universe}

In this section we present the main points of the holographic
principle, following the terminology from Ref.~\cite{3}. In the
holographic description the main component is the cut-off radius of
the horizon. According to the generalized model introduced in
Ref.~\cite{1}, the holographic energy density is inversely
proportional to the squared infrared cut-off $L_{\rm IR}$,
\begin{equation}
\rho= \frac{3c^2}{k^2L_{\rm IR}^2}, \label{1}
\end{equation}
where $k^2= 8\pi G$ is Einstein's gravitational constant; $c$ is a nondimensional and positive constant.

We will consider a homogeneous and isotropic Friedmann-Robertson-Walker (RW) metric
\begin{equation}
ds^2= -dt^2+a^2(t)\sum_{i=1}^3 (dx^i)^2, \label{2}
\end{equation}
where $a(t)$ is the scale factor.

The first Friedmann equation can be written as
\begin{equation}
H^2 = \frac{k^2}{3}\rho, \label{3}
\end{equation}
where $H(t)= \dot{a}(t)/a(t)$ is the Hubble parameter and $\rho$ is the holographic energy density.

There are several ways for choosing the infrared radius $L_{\rm IR}$: identifying with the particle horizon $L_p$,
or alternatively with the future event horizon $L_f$ \cite{2}. The definitions are
\begin{equation}
L_p(t)= a(t)\int_0^t \frac{dt'}{a(t')}, \quad L_f(t)= a(t)\int_t^\infty \frac{dt'}{a(t')}. \label{4}
\end{equation}
It is to be noted that not all choices of a cut-off can lead to an accelerating universe.

If we suppose that the energy density $\rho$ in Eq.~(\ref{3}) matches the energy density $\rho$ in Eq.~(\ref{1}),
then the first Friedmann equation for an expanding universe takes the form
\begin{equation}
H=\frac{c}{L_{\rm IR}}. \label{5}
\end{equation}
We will henceforth assume that the dark fluid, which drives the accelerated expansion, has a holographic origin. The source of this fluid can be a scalar field, or modified gravity.

\section{Holographic representation of the dissipative unified dark fluid model}

Let us consider a viscous dark fluid with an effective inhomogeneous equation of state (EoS) in flat FRW space-time   \cite{49,50},
\begin{equation}
p= \omega (\rho,t)\rho +f(\rho)-3H\zeta(H,t), \label{6}
\end{equation}
where $\omega(\rho,t)$ is the thermodynamic parameter and
$\zeta(H,t)$ is the bulk viscosity, which in general depends  on the
Hubble parameter and on the time $t$. For the function $f(\rho)$ we
choose the form  \cite{51ny}
\begin{equation}
f(\rho)= \frac{\gamma \rho^n}{1+\delta \rho^m}, \label{7}
\end{equation}
where $\gamma, \delta, n, m$ are free parameters. For thermodynamic
reasons, we take $\zeta(H,t)$ to be positive. Dissipation is
described through the bulk viscosity, for which we assume the form
\cite{48}
\begin{equation}
 \zeta(H,t)= \xi_1(t)(3H)^p, \label{8}
 \end{equation}
 with $p>0$.

 The addition of the second term in the equation of state (\ref{6}) allows us to describe the asymptotic behavior
 between the dust in the early universe and the late universe \cite{52}
 via an interpolation between different powers in the energy density expression.

 Further, we introduce nondimensional parameters
 \begin{equation}
 \tilde{\rho}= \delta^{\frac{1}{m}}\rho, \quad \tau = \frac{\gamma}{\delta^{\frac{2n-1}{2m}}}t. \label{9}
\end{equation}
Let us assume that the universe is filled by a one-component viscous fluid, and write the energy conservation law as
\begin{equation}
\dot{\rho}+3H(\rho+p)=0. \label{10}
\end{equation}
We will distinguish between two cases.

\bigskip

\noindent {\it Case 1.~}
At first , we put $\omega=-1$ in the EoS and take $\zeta (H,t)=\zeta_0$, a constant. If we restrict ourselves to $n>0$ and
$m>0$, then, provided $n-m=\frac{1}{2}$ for large $\tilde{\rho}$, Eq.~(\ref{10}) tends asymptotically to the following form,
\begin{equation}
\frac{d\tilde{\rho}}{d\tau}
\approx \sqrt{3}({\tilde{\zeta}}_0-1)\tilde{\rho}, \label{11}
\end{equation}
where  ${\tilde{\zeta}}_0 = \sqrt{3}\gamma^{-1}\delta^{\frac{2n-1}{m}}\zeta_0$. The solution of Eq.~(\ref{11}) is
\begin{equation}
\tilde{\rho} \approx {\tilde{\rho}}_0 e^{\lambda t}, \label{12}
\end{equation}
where
$\lambda = \sqrt{3}{(\tilde{\zeta}}_0-1) \frac{\gamma}{\delta^{\frac{2n-1}{m}}} $, and ${\tilde{\rho}}_0$ is an arbitrary constant.

Correspondingly, the Hubble parameter takes the form
\begin{equation}
H(t) \approx \frac{k{\tilde{\rho}}_0}{{\sqrt{3}}\,\delta^{\frac{1}{2m}}}
e^{\frac{1}{2}\lambda t}. \label{13}
\end{equation}
If $t \rightarrow 0$, the Hubble parameter approaches a constant, $H(t) \rightarrow \frac{k}{\sqrt{3}\, \delta^{\frac{1}{2m}}}.$ This case can be identified with the inflation. The scale factor is given by
\begin{equation}
a(t)= a_0\exp{\left[ \tilde{\lambda}e^{\frac{1}{2}\lambda t}\right]}, \label{14}
\end{equation}
where $\tilde{\lambda}= \frac{k{\tilde{\rho}}_0}{\sqrt{3}\, \delta^{1/2m}}$ and $a_0$ is a positive constant.

We can now calculate the particle horizon $L_p$,
\begin{equation}
L_p= \frac{2}{\lambda}\exp{\left[ \tilde{\lambda}e^{\frac{1}{2}\lambda t}\right]}
\left[ Ei\left( -\tilde{\lambda}e^{\frac{1}{2}\lambda t}\right)-Ei(- {\tilde \lambda})\right],\label{15}
\end{equation}
where $Ei(x)$ is the integral exponential function.

The Hubble parameter $H$ can be expressed in terms of the particle horizon and its time derivative as
\begin{equation}
H= \frac{{\dot{L}}_p-1}{L_p}, \quad \dot{H}= \frac{{\ddot{L}}_p}{ L_p}- \frac{{\dot{L}}_p^2}{L_p^2} + \frac{{\dot{L}}_p}{L_p^2}. \label{16}
\end{equation}
Thus by using (\ref{15}) the energy conservation equation (\ref{10}) can be rewritten as
\begin{equation}
 \frac{{\ddot{L}}_p}{ L_p}- \frac{{\dot{L}}_p^2}{L_p^2} + \frac{{\dot{L}}_p}{L_p^2} \approx \frac{\sqrt{3}\,\gamma}{2\delta^2}({\tilde{\zeta}}_0 -1)\frac{{\dot{L}}_p -1}{L_p}. \label{17}
 \end{equation}
 Thus, we have successfully applied the holographic principle to this model.

 \bigskip

 \noindent {\it Case 2.~} Let us now consider a cosmological model with a thermodynamic parameter linearly dependent on time,
 \begin{equation}
\omega(\rho, t)=at+b, \label{18}
\end{equation}
where $a,b$ are arbitrary parameters. Assume also the function $\xi_1(t)$ in Eq.~(\ref{8}) to be linear in time,
\begin{equation}
\xi_1(t)= \tau (dt+e), \label{19}
\end{equation}
with arbitrary parameters $\tau,d,e$. Then, in the case $p=1$ the bulk viscosity gets the following simple form,
\begin{equation}
\zeta(H,t)=3\tau (dt+e)H, \label{20}
\end{equation}
and the EoS gets the form
\begin{equation}
p= (at+b)\rho + \frac{\gamma \rho^n}{1+\delta \rho^m} -9\tau (dt+e)H^2. \label{21}
\end{equation}
If we put $n=m+1$, then by inserting (\ref{21}) into (\ref{10}) we obtain the modified gravitational equation of motion,
\begin{equation}
\dot{\rho}+\sqrt{3}\, k(at+b)\rho^{3/2}+\sqrt{3}\,k\rho^{1/2}\frac{\gamma \rho^{m+1}}{1+\delta \rho^m}-3\sqrt{3}\,\tau k^3(pt+e)\rho^{3/2}=0. \label{22}
\end{equation}
If   we assume $m=\frac{1}{2}$, the solution of this equation is
\begin{equation}
\rho(t)= \left[ \sqrt{ \sqrt{3}\,k\left( \frac{1}{2}a_1t^2+b_1t+\rho_0\right)}-\delta \right]^{-2}, \label{23}
\end{equation}
where $\rho_0$ is an arbitrary constant. We have here introduced the constants $a_1=\delta c_1, c_1= a-3\tau k^2d, b_1=\delta (b+1)-3\tau \delta k^2e+\gamma$, and $d_1=b+1-3\tau k^2e$.

The Hubble parameter is
\begin{equation}
H(t)= \frac{k}{\sqrt{3}}\left[ \sqrt{ \sqrt{3}\,k\left( \frac{1}{2}a_1t^2+b_1t+\rho_0\right)}-\delta \right]^{-1}. \label{24}
\end{equation}
Then in the early universe $t\rightarrow 0$, the Hubble parameter tends to the constant $H(t) \rightarrow \frac{k}{\sqrt{3}}(\sqrt{3k\rho_0}-\delta)^{-1}$, corresponding to inflaton, while in the late universe $t\rightarrow +\infty$ the Hubble parameter goes again to the constant $ H\rightarrow 0$ and we have late-time accelerating expansion

Let us put $a= 3\tau k^2d, b= 3\tau k^2e-1$, whereby $a_1=c_1=d_1=0$ and $b_1=\gamma$. In this case the Hubble parameters simplifies to
\begin{equation}
H(t)= \frac{k}{\sqrt{3}}\left[ \sqrt{\sqrt{3}\,k(\gamma t+\rho_0)}-\delta \right]^{-1}. \label{25}
\end{equation}
Let us put the constant $\rho_0=0$ and calculate the scale factor
\begin{equation}
a(t)= a_0\exp{ \left[ \frac{2k}{\sqrt{3}\,\alpha}\sqrt{t}\right]}\left( \sqrt{t}-\frac{\delta}{\alpha}\right)^{\frac{2k\delta}{\sqrt{3}\,\alpha^2}}, \label{26}
\end{equation}
where $\alpha= \sqrt{\sqrt{3}\,k\gamma}$, and calculate the particle horizon $L_p$,
\begin{equation}
L_p(t)= \exp{ \left[ \frac{2k}{\sqrt{3}\,\alpha}\sqrt{t}\right]}\left( \sqrt{t}-\frac{\delta}{\alpha}\right)^{\frac{2k\delta}{\sqrt{3}\,\alpha^2}}
\int_0^t \exp{ \left[ \frac{2k}{\sqrt{3}\,\alpha}\sqrt{t}\right]}\frac{dt}{\left( \sqrt{t}-\frac{\delta}{\alpha}\right)^{\frac{2k\delta}{\sqrt{3}\,\alpha^2}}}. \label{27}
\end{equation}
If $\delta= \frac{\sqrt{3}\,\alpha^2}{2k}$, we obtain

\begin{equation}
L_p(t)= \sqrt{3}\frac{\alpha}{k} \exp{ \left[ \frac{2k}{\sqrt{3}\,\alpha}\sqrt{t}\right]} \left( \sqrt{t}-\frac{\delta}{\alpha}\right)
\left\{  1-   \exp{ \left[- \frac{2k}{\sqrt{3}\,\alpha}\sqrt{t}\right]}     +e^{-1}\left[ Ei\left(1-\frac{2k}{\sqrt{3}\,\alpha}\sqrt{t}\right) -Ei(1)\right]      \right\}. \label{28}
\end{equation}
In this case the energy conservation equation takes the form
\begin{equation}
2\left(\frac{{\ddot{L}}_p}{ L_p}- \frac{{\dot{L}}_p^2}{L_p^2} + \frac{{\dot{L}}_p}{L_p^2}\right) +
\frac{3\sqrt{3}\,\gamma}{k}\frac{\left( \frac{{\dot{L}}_p -1}{L_p}\right)^3}
{1+\frac{\sqrt{3}\, \delta}{k}\left( \frac{{\dot{L}}_p -1}{L_p}\right)}  =0. \label{29}
\end{equation}
Thus, we have obtained a reconstruction of the conservation equation for energy, according to the holographic principle.

\section{Conclusion}

We have considered the holographic description of a unified model of the early and the late-time universe,
in a homogeneous and isotropic Friedmann-Robertson-Walker metric.
To obtain this, we have identified the infrared radius $L_{\rm IR}$ with the particle horizon $L_p$.
As a model for the universe, we have studied a general equation of state for the dark fluid in the presence of a bulk viscosity. We have explored the holographic principle for cosmological models with various values for the thermodynamic parameter $\omega (\rho,t)$ and for different forms of the bulk viscosity $\zeta(H,t)$. For each model the infrared radius, in the form of a particle horizon, has been calculated in order to obtain the energy conservation law. Thus, we have shown the equivalence between viscous models and the holographic model.

The agreement between a theoretical model of dark energy with
astronomical observations was discussed in Ref.~\cite{53}. For the
redshift parameters of distant supernova of type Ia, agreement
between observed data and theoretical prediction was obtained.

\section*{Acknowledgment}

This work was supported in part by Ministry of Education of Russian Federation, Project No FEWF-2020-0003 (A. V. T.).

\end{document}